\documentclass[journal, twocolumn]{IEEEtran}
\usepackage{cite}
\usepackage{graphicx}
\usepackage[cmex10]{amsmath}
\usepackage{color,soul}
\usepackage{setspace}
\usepackage{amsthm,amssymb}
\usepackage{mathrsfs} 
\usepackage{algorithm,algorithmic}

\begin{document}
\title{5 Gbps Optical Wireless Communication Using Commercial SPAD Array Receivers}
\author{Shenjie Huang, Cheng Chen, Rui Bian, Harald Haas, \IEEEmembership{Fellow, IEEE}, and Majid Safari, \IEEEmembership{Senior Member, IEEE}
	\thanks{Shenjie Huang and Majid Safari are with the School of Engineering, the University of Edinburgh, Edinburgh EH9 3JL, U.K. (e-mail: \{shenjie.huang, majid.safari\}@ed.ac.uk). \newline
	\hspace*{1em}Cheng Chen and Harald Haas are with LiFi Research and Development Centre, University of Strathclyde, Glasgow G1 1RD, U.K. (e-mail: \{c.chen, harald.haas\}@strath.ac.uk)\newline
	\hspace*{1em}Rui Bian is with pureLiFi, Edinburgh EH12 5EZ, U.K. (e-mail: rui.bian@purelifi.com).}
}
\maketitle
\begin{abstract}
Photon counting detectors such as single-photon
avalanche diode (SPAD) arrays can be utilized to improve the sensitivity of optical wireless communication (OWC) systems. However, the achievable data rate of SPAD-based OWC systems is strongly limited by the nonlinearity induced by SPAD dead time. In this work, the performance of SPAD-based OWC system with orthogonal frequency division multiplexing (OFDM) is investigated and compared with that of on-off keying (OOK). We employ nonlinear equalization, peak-to-average power ratio optimization by adjusting the OFDM clipping level, and adaptive bit and energy loading to achieve a record experimental data rate of $5$ Gbps. The contrasting optimal regimes of operation of the two modulation schemes are also demonstrated. 
\end{abstract}

\vspace{0.5cm}
The performance of OWC systems can be significantly degraded by occasional outages caused by low received optical power  due to channel impairments. At a low-power regime, receiver's sensitivity becomes crucial and the use of photon counting detectors such as single-photon avalanche diodes (SPAD) provides a significant advantage compared to their linear counterparts, e.g., P-i-N  photodiode (PIN PD) and avalanche photodiode (APD) \cite{Zimmermann}. 
However, when an avalanche is triggered by an incident  photon, the SPAD goes through a short period of time called \textit{dead time} during which it becomes blind to any new incident photons due to quenching process. It is known that the SPAD's dead time generates nonlinear effects including a maximum power detection limit due to saturation \cite{HuangHybrid} and a bandwidth limit due to dead-time induced intersymbol interference \cite{HuangJLT}, both of which limit the achievable data rate. In the literature, many works focused on improving the data rate of SPAD-based receivers. The current highest data rate of SPAD-based OWC system is $3.45$ Gbps as reported in \cite{Matthews21}  where OOK modulation and decision feedback (DFE) equalizer are adopted. {To deal with  the limited bandwidth of the SPAD receiver and improve its spectral efficiency beyond OOK, optical OFDM with QAM modulation has also been employed in SPAD-based systems \cite{Kosman19,Zhang22}. However, surprisingly the corresponding reported data rates, e.g., $350$ Mbps in \cite{Kosman19} and $1.35$ Gbps in \cite{Zhang22}, are far below that achieved by OOK,} which can be attributed to the limited signal's dynamic range that strongly degrades the performance of OFDM. This observation may suggest that the effect of detection power limit on the performance of SPAD receivers is the dominant factor compared to SPAD's bandwidth limit leading to the superiority of OOK.  However, in this work, we  experimentally demonstrate that a well-designed QAM-OFDM scheme adapted to SPAD nonlinearity can outperform OOK at some power levels and achieve a data rate of  $5$ Gbps which is significantly higher than the highest data rate achieved through OOK. {This record data rate is achieved by the proposed system with the mitigation of SPAD nonlinearity through joint optimization of nonlinear equalizer based on Volterra series model, peak-to-average power ratio (PAPR) adjustment by controlling OFDM clipping level, and finally the adaptive bit and energy loading algorithm.} 

\begin{figure}[!t]
	\centering\includegraphics[width=0.5\textwidth]{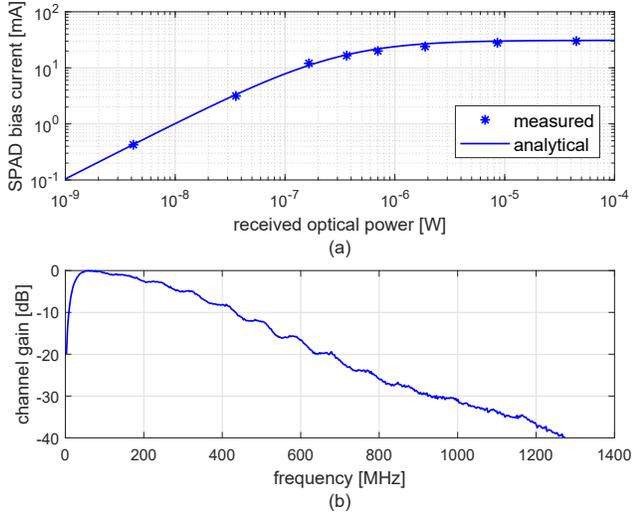}
	\caption{{(a) SPAD bias current versus the received optical power; (b) frequency response of the SPAD receiver.}}
	\label{SPAD_current_optical_power}
\end{figure}

The commercial SPAD receiver (also known as silicon photomultiplier, SiPM) employed in this work is On Semicondutor J-30020 with $N_\mathrm{SPAD}=14410$ SPAD pixels or microcells. The total area of the detector is $3.07\times3.07$ $\mathrm{mm}^2$ and the fill factor is $62 \%$ \cite{ON}. The detected photon rate of the employed SPAD receiver can be approximated as  $C={P_R\Upsilon_\mathrm{PDE}}/\left({h\nu+P_R\Upsilon_\mathrm{PDE}T_d/N_\mathrm{SPAD}}\right)$,     
where $P_R$ denotes the received signal power, $\Upsilon_\mathrm{PDE}$ is the photon detection efficiency (PDE), $h$ is Planck's constant, $\nu$ is light frequency, and $T_d$ is the dead time which is around $66$ ns \cite{MatthewsISI}. It is presented in this equation that the detected photon rate of the employed SPAD receiver has an approximately linear relationship with incident photon rate (or equivalently $P_R$) in low incident power regime, but as the incident power increases, the photocurrent becomes nonlinear and finally saturates at a fixed value $C_\mathrm{max}=N_\mathrm{SPAD}/T_d$ \cite{Matthews21}. Note that such nonlinearity is induced by the dead time and for an ideal photon counting receiver without dead time the nonlinearity effect vanishes.   
After a microcell detects a photon, it has to be recharged during the dead time. Denoting $Q$ as the charge needed for the microcell to get back to its original voltage, the current needed to maintain the operational bias voltage of the receiver is given by \cite{Matthews21}
\begin{equation}\label{Ibias}
	I_\mathrm{bias}=QC=\frac{QP_R\Upsilon_\mathrm{PDE}/h\nu}{1+P_R\Upsilon_\mathrm{PDE}T_d/N_\mathrm{SPAD}h\nu}. 
\end{equation}
As this bias current is proportional to the detected photon rate, one can use it to evaluate the saturation effect of the SPAD receiver. Fig.  \ref{SPAD_current_optical_power}(a) presents the measured bias current versus the received optical power when a bias voltage of $27.5$ V is employed. It is demonstrated that the measured bias current matches with that of the analytical expression given in (\ref{Ibias}). Note that the parameters $Q=0.14$ pC and $\Upsilon_\mathrm{PDE}=0.36$ are employed to achieve the analytical result which are close to the values selected in \cite{Matthews21}. The nonlinear response of the SPAD receiver can be clearly observed in Fig.  \ref{SPAD_current_optical_power}(a). Such nonlinearity can strongly limit the performance of SPAD receiver in high power regimes. In this work, we utilize the fast output of the employed commercial SPAD receiver  \cite{Matthews21}. The measured 3-dB bandwidth of the receiver is around $250$ MHz as illustrated in Fig.  \ref{SPAD_current_optical_power}(b). 
   
\begin{figure}[!t]
	\centering\includegraphics[width=0.47\textwidth]{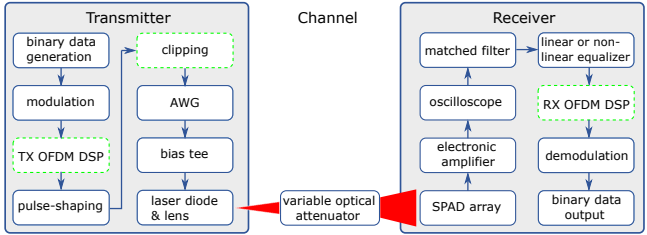}
	\caption{{The schematic diagram of the experimental setup. The dashed blocks are additional steps when OFDM is employed.}} 
	\label{system_model}
\end{figure}

Fig. \ref{system_model} presents the block diagram of the experimental setup. The demonstration is operated in dark condition. At the transmitter side, the binary data stream is firstly generated using MATLAB in PC. After applying the modulation (OOK or DC-biased optical OFDM) and root-raised-cosine (RRC) pulse-shaping, the modulated discrete signal with desired sampling rate is sent to arbitrary waveform generator (AWG, Keysight M8195A) to be converted to the analog waveform. {When OFDM is considered, extra steps, i.e., transmitter-side OFDM DSP and signal clipping, should be employed. The former mainly contains IFFT operation and adding cyclic prefix \cite{Rui19}.}      
The output AC signal of the AWG is then combined with the DC bias from a power supply (Keysight E36313A) by utilizing a bias tee (Mini-Circuits ZFBT-6GW+). The output signal of the bias tee is used to drive the laser diode (Thorlabs L405P20) which has a central wavelength of $405$ nm and a 3-dB bandwidth above $800$ MHz. This laser diode is employed as its central wavelength is close to the peak PDE wavelength of the SPAD receiver, i.e., $420$ nm. {In the demonstration, the bias voltage of the laser diode and the Vpp of the modulated electrical signal are set to $4.55$ V and $0.8$ V, respectively, to fully utilize the laser linear dynamic range.} Finally, an aspheric lens is used to collimate the laser beam. 

For OWC systems, the received optical power normally changes over time due to various channel effects, e.g., scintillation in FSO and random orientation in VLC. In order to emulate the change of optical power introduced by practical OWC channels, a variable optical attenuator is used which comprises a neutral density filter wheel (Thorlabs FW1A) and a wire grid polarizing film (Edmund optics 34254).  
By rotating the filter wheel and the polarizer, the continuous change of the channel path loss and hence the received signal power can be achieved.     
The laser beam after propagating through the attenuator illuminates the SPAD array receiver.
The laser beam is  diffracted to avoid the excessive received optical power which might damage the SPAD array. The output of the SPAD receiver is amplified by the electronic amplifier (Mini-Circuits ZX60-43-S+) and then fed to the oscilloscope (Keysight DSA90804A) which has a bandwidth of $8$ GHz. The received waveform captured by the oscilloscope is sent to the PC for the offline processing. The matched filtering which matches with the signal pulse is applied and the output downsampled signal then goes through specific equalizer which varies with the investigated system. {Additional receiver-side OFDM DSP should be employed when OFDM is considered.} Finally, the signal decoding process is adopted to recover the transmitted information bits.   

In this work, we consider two commonly used modulation schemes, i.e., OOK and  DCO-OFDM. For the system with OOK, two different equalization methods are considered at the receiver, i.e., linear and nonlinear equalization. In both equalization methods, the recursive least square (RLS) algorithm is utilized to train the tap weights. As the dead-time-induced effects are inherently nonlinear, it is expected that nonlinear equalization (NEQ) can result in superior performance compared to that of linear equalization (LEQ). The employed NEQ is designed based on Volterra series model. Considering the complexity of the higher-order Volterra filter, only the first two orders of the Volterra series are considered \cite{Zhou:20}. The output of NEQ can be expressed as
$d(n)=$
\begin{equation}\label{NEQ}
	\!\!\!\sum_{i=0}^{N_l-1}\!\!\! w_1(i)x(n-i)+\!\!\!\sum_{m_1=0}^{N_{nl}-1}\!\sum_{m_2=m_1}^{N_{nl}-1}\!\!\!\!\!w_{2}(m_1,m_2)x(n-m_1)x(n-m_2),
\end{equation} 
where $x(n)$ denotes the signal before equalization, $N_l$ and $N_{nl}$ refer to the memory length of the linear ($1$st order) and nonlinear (2nd order) terms, respectively, and $w_1$ and $w_2$ are the corresponding tap weights.  In this demonstration, in order to balance the complexity and performance, the memory lengths of  NEQ are set to  $N_l=41$ and $N_{nl}=21$, respectively.     

\begin{figure}[!t]
	\centering\includegraphics[width=0.48\textwidth]{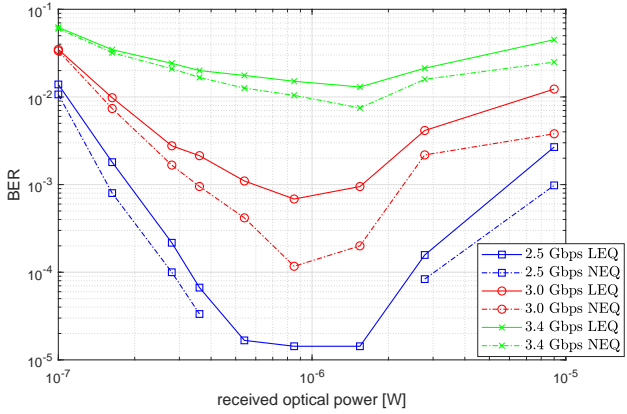}
	\caption{The BER versus the received optical power under various data rates in the presence of LEQ or NEQ when OOK modulation is employed.}
	\label{OOK_BER}
\end{figure}
\begin{figure}[!t]
	\centering\includegraphics[width=0.49\textwidth]{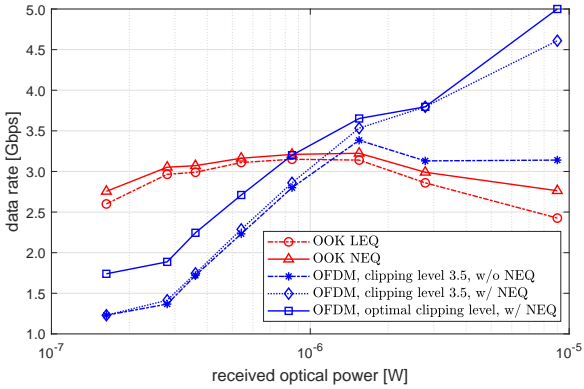}
	\caption{The achievable data rate versus the received signal power for the considered modulation schemes. }
	\label{DR}
\end{figure}

Fig. \ref{OOK_BER} presents the measured BER versus the received optical power when OOK modulation is employed. It is demonstrated that with the increase of the received optical power the BER firstly decreases and then increases. 
This is because in low received power regime, the system is signal power limited and hence increase the signal power can improve the BER performance; however, with a further increase of signal power, the dead-time-induced nonlinearity come into effect which in turn degrade the performance. Similar BER behaviour has also been reported in many theoretical works \cite{Khalighi192,HuangJLT,HuangHybrid}. {The superiority of NEQ over its linear counterpart can also be clearly observed. However, it is demonstrated that with the increase of the data rate, the improvement of NEQ drops.
This attributes to the fact that the increase of data rate means higher signalling bandwidth which greatly increases the effective noise power. Therefore, for high data rate scenario, the performance is more dominated by the noise power and the improvement introduced by NEQ becomes less.}
\begin{figure}[!t]
	\centering\includegraphics[width=0.48\textwidth]{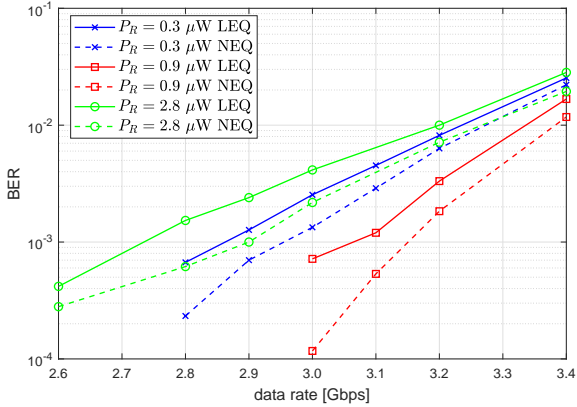}
	\caption{The BER versus the data rate under various received optical power $P_R$ when OOK modulation is employed. }
	\label{OOK_BER_datarate}
\end{figure}
Fig. \ref{OOK_BER_datarate} presents the BER versus the data rate under various received optical power. One can observe that for a specific BER target, the transmission with NEQ can provide higher data rate than that with LEQ. 
Moreover, it is again presented that larger $P_R$ does not necessarily result in better performance due to the SPAD nonlinearity. Based on Fig. \ref{OOK_BER_datarate}, for any BER target the achievable data rate versus the received power results can also be achieved, as presented later in Fig. \ref{DR}.

{In general, OFDM with QAM modulation can be employed to improve the spectral efficiency.} Two different DCO-OFDM-based SPAD systems are investigated in this work. The first is the traditional one with single-tap frequency domain equalization \cite{Chen16}. For the second system, the time-domain NEQ shown in (\ref{NEQ}) is further employed to mitigate the nonlinear effects. The NEQ at the receiver is applied to the signal after matched filtering and before the Fourier transform operation. The same memory lengths of NEQ as OOK transmission are employed to ensure the fair comparison between the two modulation schemes.  We consider that for both OFDM systems the modulation bandwidth is $1400$ MHz and the number of subcarriers is $1024$. OFDM enables the use of the adaptive bit and energy loading algorithm to improve the achievable data rate \cite{Rui19}, which is also adopted in the considered systems.  

The time-domain OFDM signal is approximately normal distributed which has a high PAPR; however, in practical communication systems with limited dynamic range, this will lead to a significant decrease in transmitted signal power. Therefore, signal clipping is crucial for OFDM transmission. In the proposed systems, a normalized bipolar OFDM signal in time domain which follows a normal distribution with zero mean and unity variance is firstly generated. The signal is then clipped so that the maximal and minimal amplitude are $\varepsilon$ and $-\varepsilon$, respectively, where $\varepsilon$ denotes the normalised clipping level. Later, the normalized electrical signal is scaled to a desired Vpp to drive the laser diode, which is $0.8$ V in the proposed systems. The change of $\varepsilon$ is equivalent to the change of the PAPR of the transmitted OFDM signal. By selecting larger $\varepsilon$, it is expected that the received optical signal can occupy less dynamic range of the receiver and hence the effect of SPAD nonlinear distortion can be reduced. In addition, choosing larger clipping level can reduce the clipping distortion. However, since the Vpp of the transmitted signal is fixed, larger clipping level also results in less signal power which in turn degrades the performance \cite{Chen16}. Therefore, it is expected that the clipping level, or equivalently the PAPR, of the transmitted OFDM signal can be optimized to achieve the best performance.

\begin{figure}[!t]
	\centering\includegraphics[width=0.5\textwidth]{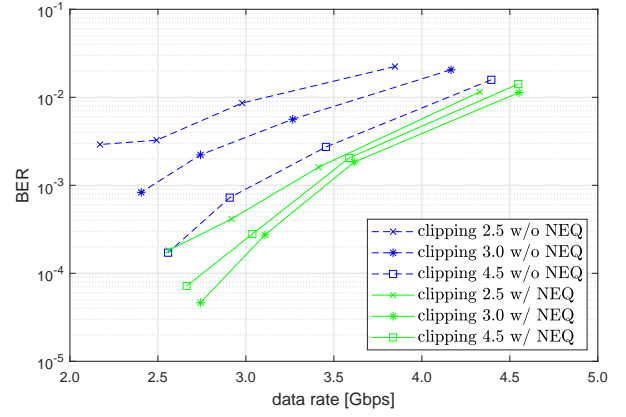}
	\caption{The BER of OFDM versus the data rate under various clipping levels when the received signal power is $2.8$ $\mu$W.} 
	\label{OFDM_BER_datarate_28mA_NEQ}
\end{figure}
\begin{figure}[!t]
	\centering\includegraphics[width=0.48\textwidth]{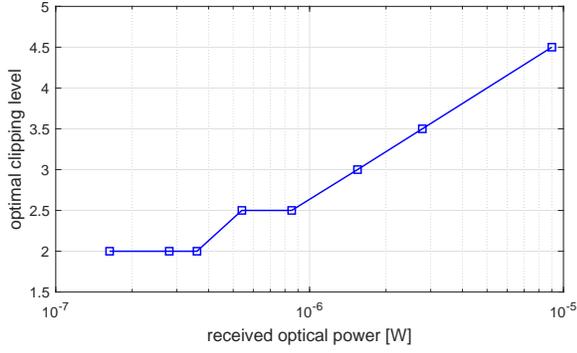}
	\caption{The optimal clipping level for OFDM versus the received signal power in the presence of NEQ. }
	\label{clipping}
\end{figure}
Fig. \ref{OFDM_BER_datarate_28mA_NEQ} presents the BER of the OFDM transmission versus the data rate with various normalized clipping levels. The superiority of the system with NEQ over that without NEQ can be clearly observed. In addition, it is demonstrated that the system performance varies with the employed clipping level due to aforementioned trade-off.
{Because of the stronger encountered nonlinear effects, system without NEQ benefits more from the increase of the clipping level. Therefore, different from system with NEQ, the degradation of performance does not appear even when the largest considered clipping level $4.5$ is employed, which means that the optimal clipping level for such system could be further above this level.}
Fig. \ref{OFDM_BER_datarate_28mA_NEQ} also indicates that in order to achieve the highest achievable data rate, one can employ both the NEQ and clipping optimization. The measured optimal clipping level versus the received optical power when NEQ is employed is presented in Fig. \ref{clipping}. To achieve this result, a vector of clipping levels ranging from $2$ to $4.5$ with a step size of $0.5$ is considered. For each received optical power, the clipping level which provides the highest achievable data rate is recorded. It is illustrated that with the increase of the received signal power, the optimal clipping level also increases to account for higher nonlinear effect. 

Finally, Fig. \ref{DR} demonstrates the measured achievable data rate versus the received optical power for both OOK and OFDM with a BER target of $2\times 10^{-3}$ which is below the $7\%$ FEC limit. For OOK transmission, with the increase of the optical power, the  increase and then decrease of the data rate can be observed. 
It is illustrated that employing NEQ can slightly improve the date rate. The measured highest achievable data rates of OOK are $3.15$ Gbps and $3.22$ Gbps when linear and nonlinear equalizers are employed, respectively, which are achieved at a received power of $0.85$ $\mu$W.  This result is close to the record data rate of SPAD-based system with OOK, i.e., $3.6$ Gbps, reported in \cite{Matthews21}. To achieve a data rate of $3$ Gbps, the required received optical power is $0.28$ $\mu$W which corresponds to a sensitivity of $-35$ dBm. Considering that corresponding sensitivities of the state-of-the-art PIN PD and APD receivers are around $-20$ dBm and $-30$ dBm \cite{Juki}, the superiority of SPAD receiver over its counterparts in terms of the sensitivity is demonstrated.       
\begin{figure}[!t]
	\centering\includegraphics[width=0.48\textwidth]{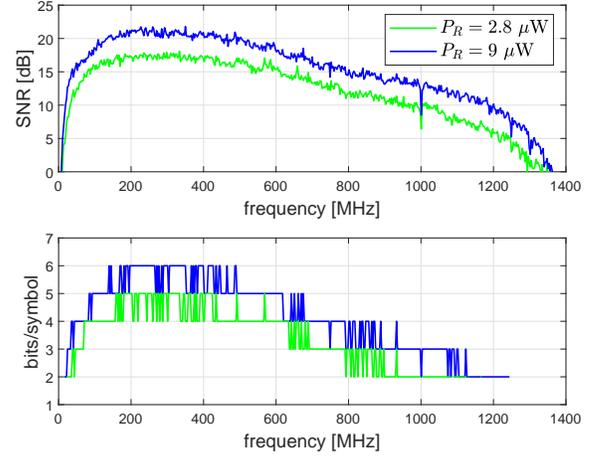}
	\caption{The measured SNR and adaptively loaded bits for each subcarrier of OFDM when NEQ and the optimal clipping level are employed.} 
	\label{OFDM_SNR_bits}
\end{figure}
On the other hand, when OFDM modulation is employed, in the absence of clipping optimization and NEQ, the change of the data rate with the increase of optical power is similar to that when OOK is employed. However, different from OOK transmission, employing NEQ in OFDM can significantly improve the data rate. For instance, when the received power is $9$ $\mu$W and the clipping level is $3.5$, the achievable data rate is $3.14$ Gbps in the absence of NEQ; however, when NEQ is employed, this data rate rises to $4.6$ Gbps. {This is because for OFDM the data is modulated on the waveform rather than two amplitude levels as OOK and hence OFDM is more vulnerable to nonlinearity which can be effectively mitigated through NEQ.} It is also presented in Fig. \ref{DR} that the clipping level optimization can further improve the data rate performance. By employing both NEQ and clipping level optimization, achievable data rates of $3.8$ Gbps and $5$ Gbps have been achieved when the received power is $2.8$ $\mu$W and $9$ $\mu$W, respectively. The measured SNR and bit loading results for these two cases are shown in Fig. \ref{OFDM_SNR_bits}. It is illustrated that increasing the signal power from $2.8$ $\mu$W to $9$ $\mu$W can introduce around $3$ dB SNR improvement over the whole considered spectrum range which strongly improves the spectrum efficiency. 

The comparison of the achievable data rates of OOK and OFDM transmission presented in Fig. \ref{DR} further indicates that these two modulation schemes have contrasting optimal regimes of operation.  {When the received power is relatively low, due to the low dynamic range of the received optical signal, OOK can provide significantly higher data rates than OFDM which requires wide dynamic range to achieve decent performance due to the high PAPR of its time-domain signals.} However, OFDM in turn outperforms OOK in high power regime. Therefore, in practical implementation, OOK is preferable for applications with lower received power; whereas, for applications with higher received power, OFDM is a preferred choice. {For links with  highly dynamic channels, the adaptive switching between OOK and OFDM based on the instantaneous channel condition may be employed to achieve the best  performance over a wide range of received power.}

\section*{Acknowledgements}
We gratefully acknowledge the financial support from EPSRC under grant EP/R023123/1 (ARROW) and EP/S016570/1 (TOWS). For the purpose of open access, the author has applied a Creative Commons Attribution (CC BY) licence to any Author Accepted Manuscript version arising from this submission.

\bibliographystyle{IEEEtran}
\bibliography{IEEEabrv,manuscript}
\end{document}